\documentclass[pra,aps,amsmath,amssymb,amsfonts,twocolumn,nofootinbib,longbibliography,superscriptaddress]{revtex4-2}
\usepackage{amsfonts}
\usepackage{txfonts}
\usepackage{amsmath}
\usepackage{float}
\usepackage{diagbox}
\usepackage[colorlinks,breaklinks,linkcolor=blue,anchorcolor=blue,citecolor=blue,urlcolor=blue]{hyperref}
\usepackage{graphicx}
\usepackage{dcolumn}
\usepackage{bm}
\usepackage{amssymb}
\usepackage{mathrsfs}
\usepackage[sort&compress]{natbib}
\usepackage{subfigure}
\usepackage{braket}
\usepackage{xcolor}

\usepackage{lineno}

\begin{document}
\title{Giant-Atom-Induced High-Order Output Zeros in a Coupled-Cavity Array}

\author{Mengxue Li}
\affiliation{Center for Quantum Sciences and School of Physics, Northeast Normal University, Changchun 130024, China}

\author{H. Z. Shen}
\affiliation{Center for Quantum Sciences and School of Physics, Northeast Normal University, Changchun 130024, China}

\author{H. D. Liu}
\affiliation{Center for Quantum Sciences and School of Physics, Northeast Normal University, Changchun 130024, China}

\author{Gangcheng Wang}
\email{wanggc887@nenu.edu.cn}
\affiliation{Center for Quantum Sciences and School of Physics, Northeast Normal University, Changchun 130024, China}

\date{\today}

\begin{abstract}
Coherent perfect absorption, zero transmission and zero reflection are several scattering phenomena governed by interference engineering in Hermitian and non-Hermitian systems. Higher-order coherent perfect absorption can significantly broaden the absorption bandwidth, while existing implementations rely on scattering zero degeneracy induced by exceptional points or additional momentum-dependent phase delays introduced in incident waves. We propose a scheme where a giant atom couples to a one-dimensional coupled-cavity array at two spatially separated sites. The spatially separated coupling configuration of the giant atom generates tunable nonlocal interference phases that dominate the scattering interference process. We further investigate the zero-reflection and zero-transmission behaviors under single-sided incidence. Remarkably, we find that for certain parameter choices, zero transmission can persist over the entire propagating band, rather than being restricted to a single momentum. Our results reveal that the giant-atom interference mechanism enables bandwidth-enhanced coherent perfect absorption and bandwidth-enhanced zero transmission in the absence of exceptional points and incident momentum-dependent phase delays. Our work provides a physical route for coherent wave manipulation in coupled-cavity quantum networks.
\end{abstract}
\maketitle

\section{Introduction}

Coherent control of light–matter interactions constitutes a core research topic in modern quantum optics and photonic information processing \cite{PhysRevA.92.023824,PhysRevA.93.063805,PhysRevA.95.013841,PhysRevA.83.055804}. Among various interference-based physical phenomena, coherent perfect absorption (CPA) was originally proposed as the time-reversed counterpart of lasing \cite{PhysRevLett.105.053901,science.1200735}. Enabled by the synergy between destructive interference and loss mechanisms, CPA achieves complete suppression of the outgoing fields \cite{PhysRevLett.105.053901,PhysRevLett.108.186805}.
Furthermore, CPA has been realized across diverse platforms, including plasmonic metasurfaces \cite{Zhang2012,PhysRevB.86.165103,Dutta-Gupta:12}, graphene films \cite{Rao:14,Pirruccio2013}, optical waveguides \cite{refId0,Park:15}, and sound waves \cite{Meng2017}. Owing to its fundamental research value and promising applications in optical switches \cite{10.1063/1.4870635,6359733,Guo:23}, single-port interferometers \cite{Li2014,Jin:20}, and sensors \cite{9804238,Li2019}, CPA has attracted considerable interest \cite{PhysRevLett.108.186805,PhysRevLett.109.257402,PhysRevA.82.031801,PhysRevLett.107.033901,PhysRevLett.106.093902,PhysRevLett.107.163901,PhysRevA.82.021802,PhysRevA.93.063805,PhysRevA.95.013841,Roger2015,Zhang2017}.

Nevertheless, conventional CPA is typically a narrowband effect. Higher-order CPA is generally closely associated with exceptional points (EPs) \cite{science.abj1028,Soleymani2022,PhysRevLett.112.203901,Chen2017,Hodaei2017,PhysRevA.107.062209}, at which multiple purely incident scattering solutions coincide, giving rise to high-order output zeros \cite{PhysRevLett.122.093901,PhysRevB.95.144303}. This phenomenon has been predicted and observed in optical \cite{PhysRevLett.133.173801,science.abj1028,Soleymani2022}, electronic \cite{Suwunnarat2022}, and acoustic systems \cite{slhy-f76q}, exhibiting great potential for bandwidth-enhanced absorption. However, such EP-based mechanisms suffer from inherent limitations. High-order scattering responses rely strictly on precise parameter tuning and inherit the intrinsic spectral characteristics of EPs, which renders the system more susceptible to external perturbations \cite{Feng:24,18gg-gvzc}. Recent studies have revealed that high-order perfect absorption can also be realized by manipulating the momentum-dependent properties of incident waves without relying on EPs \cite{nkls-pgkf}.

In addition to CPA, precise manipulation of optical scattering under single-port incidence, such as perfect transmission and perfect reflection, is another essential research objective in photonic networks \cite{PhysRevA.74.043818,PhysRevLett.101.100501,PhysRevA.80.014301,PhysRevX.3.031013,Zhu_2019,PhysRevA.100.053851,PhysRevResearch.4.L032015}. In Hermitian systems, perfect transmission and reflection originate from unitary interference processes, which redistribute optical energy among different transport channels without dissipation. In non-Hermitian systems with intrinsic loss, the interplay between interference and dissipation enables the realization of zero-reflection and zero-transmission behaviors, which are essential for implementing single-photon switching. A variety of single-photon switching and routing schemes have been experimentally demonstrated in both microwave and optical frequency regimes \cite{science.1152261,PhysRevLett.102.083601,PhysRevLett.107.073601,PhysRevLett.111.193601,Papon:19}. Numerous theoretical schemes reported to date suffer from frequency sensitivity, restricting switching behaviors to an extremely narrow spectral range \cite{PhysRevA.89.013805,PhysRevA.99.033827,PhysRevA.99.063815,PhysRevLett.111.103604}. Even a small frequency detuning can substantially degrade the scattering performance, making it crucial to develop bandwidth-enhanced scattering mechanisms for both fundamental research and practical photonic device applications.

In recent years, giant atoms have emerged as a novel paradigm for manipulating light–matter interactions. Unlike conventional small atoms that couple locally to waveguides at a single position \cite{PhysRevLett.95.213001}, giant atoms interact with propagating optical fields at multiple spatially separated sites \cite{PhysRevA.90.013837}. This nonlocal coupling introduces additional interference pathways, enabling flexible and precise modulation of system scattering properties. It facilitates the exploration of diverse physical phenomena, including frequency-dependent Lamb shifts \cite{PhysRevA.90.013837,PhysRevA.95.053821}, bound states \cite{PhysRevResearch.2.043014,PhysRevLett.126.043602,PhysRevA.102.033706,PhysRevA.104.053522,Xiao2022,PhysRevA.107.023716}, decoherence-free interaction \cite{PhysRevResearch.2.043184,Kannan2020,PhysRevA.105.023712,PhysRevLett.120.140404,PhysRevResearch.2.043070}, non-Markovian decay dynamics \cite{PhysRevA.103.053701,PhysRevA.106.063703,Qiu2023,PhysRevResearch.6.033243,PhysRevA.110.033707}, relaxation rates \cite{PhysRevA.103.023710,PhysRevLett.126.043602,PhysRevLett.128.223602} and few-photon scattering \cite{PhysRevA.104.033710,PhysRevA.104.023712,PhysRevA.106.013715,PhysRevA.108.063715,PhysRevA.108.053718,PhysRevA.111.023711}. By tuning the spatial distance between coupling sites and the relative coupling phase, specific scattering channels can be selectively enhanced or suppressed \cite{Chen2022,Wang_2022,PhysRevX.13.021039,PhysRevLett.133.063603}, allowing controllable switching between zero reflection and zero transmission. Owing to these unique advantages, giant atom systems provide an ideal platform for investigating complex scattering phenomena.

In this paper, we systematically investigate the scattering characteristics of a two-site coupled giant atom waveguide system, with a particular focus on the physical conditions for achieving CPA, zero reflection and zero transmission. We derive the analytical expressions for reflection and transmission amplitudes, and demonstrate that the giant-atom configuration enables flexible modulation of the order of output zeros. Through deliberate parameter selection, we realize second-order CPA and third-order zero transmission, verifying the existence of high-order zeros in scattering spectra. The proposed mechanism may be applicable to various platforms, including coupled-cavity arrays \cite{PhysRevA.102.063709}, LC-circuit resonators \cite{l1fq-gbbl}, acoustic waveguides \cite{Manenti2017}, or transmission-line waveguides \cite{PhysRevA.103.023710,Kannan2020}. This study provides physical insight and a feasible route for designing bandwidth-enhanced photonic devices based on coherent interference manipulation.

The remainder of this paper is structured as follows. In Sec. \ref{Sec:II}, we introduce the model consisting of a two-level giant atom coupled to an infinite one-dimensional tight-binding cavity array at two spatially separated positions, and derive the corresponding scattering matrix of the system. In Sec. \ref{Sec:III}, we elaborate the physical conditions for CPA and present the parameter regimes for realizing first-order and second-order CPA. In Sec. \ref{Sec:IV}, we investigate the requirements for zero-reflection and zero-transmission behaviors and clarify the parameter conditions for output zeros of different orders. In Sec. \ref{Sec:V}, we perform wave-packet time-evolution simulations to validate our theoretical analyses. In Sec. \ref{Sec:VI}, we present the experimental implementation scheme and verify the feasibility of the proposed system. Finally, we summarize the main results in Sec. \ref{Sec:VII}.

\section{Model and Scattering Formalism}
\label{Sec:II}

\subsection{Model and single-excitation dynamics}

\begin{figure}
\centering
\includegraphics[width=0.9\columnwidth]{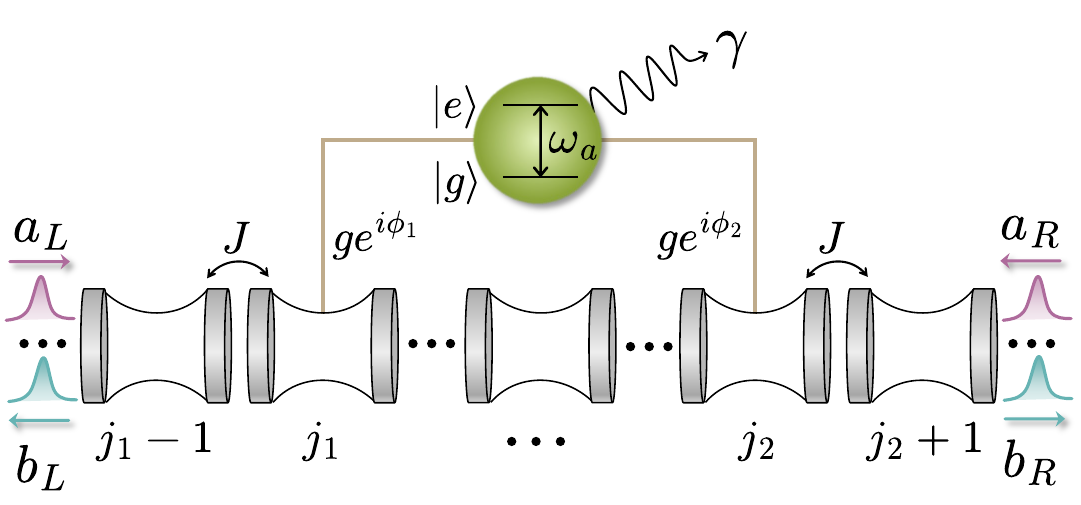}
\caption{Schematic of a two-site coupled giant atom interacting with a coupled-resonator waveguide. The waveguide consists of resonators interconnected with hopping strength $J$. The transition frequency of the two-level giant atom is $\omega_a$, and $\gamma$ stands for the atomic dissipation rate. The giant atom couples to the $j_1$th and $j_2$th resonators with coupling strengths $ge^{i\phi_1}$ and $ge^{i\phi_2}$, respectively.}
\label{Fig.1}
\end{figure}
As illustrated in Fig.~\ref{Fig.1}, we consider an infinite
one-dimensional coupled-cavity array described by the standard tight-binding model for coupled-resonator waveguides
\cite{Yariv1999CROW}, interacting with a two-level giant atom at two spatially separated lattice sites $j_1$ and $j_2$. The separation between the two coupling points is $L=j_2-j_1$. For an infinite one-dimensional cavity array, the physical scattering and absorption properties depend on the relative separation between the two coupling points rather than on the global position of the atom. Therefore, we are free to set the origin of the coordinate system at the midpoint of the two coupling sites (i.e., $j_{1}=-L/2$ and $j_{2}=L/2$). For odd $L$, the midpoint convention shifts all lattice coordinates by half a lattice spacing. Throughout this work, we set $\hbar=1$ and take the lattice constant as the unit of length. The total effective Hamiltonian is
\begin{equation}\label{eq_H}
\hat{H}=\hat{H}_c+\hat{H}_a+\hat{H}_{\rm int},
\end{equation}
where the Hamiltonian for the cavity array and the dissipative giant atom is given by
\begin{equation}
\begin{aligned}
\hat{H}_c &= \omega_c\sum_j \hat{a}_j^\dagger \hat{a}_j - J\sum_j \left( \hat{a}_{j+1}^\dagger \hat{a}_j+\hat{a}_j^\dagger \hat{a}_{j+1} \right),\\
\hat{H}_a &= \left(\omega_a-\frac{i\gamma}{2}\right)\hat{\sigma}^+\hat{\sigma}^-,
\end{aligned}
\end{equation}
with the dissipation rate satisfying $\gamma\geq 0$. The nonlocal giant-atom and coupled-cavity array interaction is described by
\begin{equation}
\hat H_{\rm int}
=
\sum_{\mu=1}^{2}
\left(
G_\mu \hat a_{j_\mu}^\dagger\hat\sigma^-
+
G_\mu^*\hat a_{j_\mu}\hat\sigma^+
\right),
\label{eq:Hint}
\end{equation}
with $G_\mu=g_\mu e^{i\phi_\mu}$. The phases $\phi_\mu$ are local coupling phases and may, for example,
be controlled by external magnetic fluxes in superconducting
implementations \cite{Wang_2022,Chen2022}. We retain the general coupling magnitudes $g_1$ and
$g_2$ in the derivation below and subsequently specialize to the symmetric case $g_1=g_2 = g$ used in the main analysis.
$\hat{a}_j^\dagger$ ($\hat{a}_j$) is the creation (annihilation) operator for the $j$th cavity mode, and $\hat{\sigma}^+$ ($\hat{\sigma}^-$) is the raising (lowering) operator for the giant atom. $\omega_c$ and $\omega_a$ denote the frequencies of a single node of the cavity array and the transition frequency of the giant atom, respectively. The parameter $J$ characterizes the nearest-neighbor hopping rate and $\gamma$ denotes the dissipation of the atom.

Because the Hamiltonian conserves the total excitation number apart from the effective atomic loss, a stationary single-excitation state with energy $\omega(k)$ can be written as
\begin{equation}
|\Psi_{k}\rangle
=
\left(
\sum_j u_j\hat a_j^\dagger
+
u_e\hat\sigma^+
\right)|g,0\rangle,
\label{eq:single_state}
\end{equation}
where $u_j$ and $u_e$ are the probability amplitudes for finding the excitation in the $j$th cavity and in the giant atom, respectively. $|g,0\rangle$ represents that the atom is in the ground state $|g\rangle$ while no photon occupies the cavity. Substitution of Eq.~\eqref{eq:single_state} into the stationary
Schr\"odinger equation
$\hat H|\Psi(k)\rangle=\omega(k)|\Psi(k)\rangle$ gives
\begin{equation}\label{eq:lattice_equation}
\begin{aligned}
&[\omega(k)-\omega_c]u_j =
-J(u_{j+1}+u_{j-1})+\sum_{\mu=1}^{2}
G_\mu u_e\delta_{j,j_\mu},\\
&\left[\omega(k)-\omega_a+i\frac{\gamma}{2}\right]u_e =
\sum_{\mu=1}^{2}G_\mu^*u_{j_\mu}.
\end{aligned}
\end{equation}
Away from the two coupling sites, the first equation in Eq.~\eqref{eq:lattice_equation} reduces to the free tight-binding equation of a coupled-resonator waveguide
\cite{Yariv1999CROW}. A plane wave
$u_j\propto e^{ikj}$ therefore obeys $\omega(k)=\omega_c-2J\cos k$, with $0<k<\pi$. 
Its group velocity is $v_g(k)=d\omega(k)/dk=2J\sin k$. Consequently, $e^{ikj}$ and $e^{-ikj}$ represent right- and
left-propagating waves, respectively, within the wave-vector interval
considered here.

\subsection{Scattering matrix}
 
The right-moving wave on the left side $a_{L}$ and the left-moving wave on the right side $a_{R}$ constitute the two input channels, while the waves propagating in the opposite directions correspond to the two output channels $b_{L/R}(k)$. The input-output relation for the two-port scattering is governed by $\boldsymbol{b}(k)=S(k)\boldsymbol{a}$, where $\boldsymbol{b}(k)=[b_L(k),b_R(k)]^T$ and $\boldsymbol{a}=(a_L,a_R)^T$. 
The scattering matrix $S(k)$ is defined as
\begin{equation}
S(k) =\begin{bmatrix}
r_L(k) & t_R(k) \\
t_L(k) & r_R(k) 
\end{bmatrix}.
\end{equation}
Here, $r_L(k)$ and $t_L(k)$ are the reflection and transmission amplitudes
for left incidence, whereas $r_R(k)$ and $t_R(k)$ correspond to right
incidence. Because the coupling constants $G_\mu$ may carry
nontrivial phases, the scattering amplitudes associated with the two
incident directions should, in general, be retained separately.

We first consider a unit-amplitude plane wave incident from the left,
i.e., $\boldsymbol a=(1,0)^T$. The cavity amplitudes can be expressed
as
\begin{equation}\label{eq_A1}
u_j^{(\rightarrow)}=
\begin{cases}
e^{ikj}+r_L(k)e^{-ikj}, & j<j_1,\\
\alpha_\rightarrow e^{ikj}+\beta_\rightarrow e^{-ikj}, & j_1\leq j\leq j_2,\\
t_L(k) e^{ikj}, & j>j_2,
\end{cases}
\end{equation}
where $\alpha_\rightarrow$ and $\beta_\rightarrow$ represent the amplitudes between two coupling points. Substituting Eq. (\ref{eq_A1}) into the discrete Schr\"odinger equations at $j_1$ and $j_2$, together with the single-valuedness of the lattice amplitudes at the two coupling sites, gives the reflection and transmission amplitudes
\begin{equation}
r_L(k)
=
-\frac{i}{v_g(k)}V_-(k)u_e^{(\rightarrow)},
\quad
t_L(k)
=
1-\frac{i}{v_g(k)}V_+(k)u_e^{(\rightarrow)}.
\label{eq:left_rt_intermediate}
\end{equation}
We have introduced the direction-dependent collective coupling
amplitudes
\begin{equation}
V_+(k)
=
\sum_{\mu=1}^{2}G_\mu e^{-ikj_\mu},
\quad
V_-(k)
=
\sum_{\mu=1}^{2}G_\mu e^{ikj_\mu}.
\label{eq:Vpm}
\end{equation}
The quantities $V_+(k)$ and $V_-(k)$ denote the total radiation amplitudes produced by far‑field interference of photons emitted into the right‑ and left‑propagating waveguide modes from the two spatially separated coupling sites after the giant atom is excited. Their momentum dependence originates from the propagation phases accumulated between the spatially separated
coupling points. Thus, $V_\pm(k)$ contain the central interference
mechanism responsible for the scattering responses discussed below.

The amplitudes at the two coupling sites may be written compactly as
\begin{equation}
u_{j_\mu}^{(\rightarrow)}
=
e^{ikj_\mu}
-\frac{i}{v_g(k)}
\sum_{\nu=1}^{2}
G_\nu e^{ik|j_\mu-j_\nu|}
u_e^{(\rightarrow)}.
\label{eq:coupling_site_amplitudes}
\end{equation}
Inserting Eq.~\eqref{eq:coupling_site_amplitudes} into the second equation in Eq.~\eqref{eq:lattice_equation} yields
\begin{equation}
u_e^{(\rightarrow)}
=
\frac{V_+^*(k)}{D(k)},
\label{eq:ue_left}
\end{equation}
where $D(k)=\omega(k)-\omega_a-\Delta_g(k)+i\left[\gamma+\Gamma_g(k)\right]/2$ is the dressed scattering denominator. The cavity-array-induced
frequency shift and radiative decay rate are obtained as follows
\begin{equation}
    \begin{aligned}
        \Delta_g(k)&=\frac{2g^2}{v_g(k)}\cos\Delta\phi\sin  kL,\\
        \Gamma_g(k)&=\frac{4g^2}{v_g(k)}(1+\cos\Delta\phi\cos  kL),
    \end{aligned}
\end{equation}
where $\Delta\phi=\phi_2-\phi_1$. Without loss of generality, we restrict $\Delta\phi$ to the interval $(-\pi,\pi]$. The real part of $D(k)$ describes the detuning between the incident photon and the giant atom after including the cavity array induced frequency shift, where $\Delta_g(k)$ represents the array-induced shift of the atomic resonance, while the imaginary part gives the total linewidth, including both the atomic loss $\gamma$ and the radiative decay $\Gamma_g(k)$ into the cavity array. Considering Eqs.~\eqref{eq:left_rt_intermediate} and
\eqref{eq:ue_left}, we obtain
\begin{equation}
r_L(k) = -\frac{iV_-(k)V_+^*(k)}{v_g(k) D(k)},\quad t_L(k) = 1- \frac{i|V_+(k)|^2}{v_g(k) D(k)}.
\end{equation}

For completeness, we next consider a unit-amplitude wave incident
from the right, corresponding to $\boldsymbol a=(0,1)^T$. The
scattering ansatz is
\begin{equation}
u_j^{(\leftarrow)}
=
\begin{cases}
t_R(k)e^{-ikj}, & j<j_1,\\[2mm]
\alpha_\leftarrow e^{ikj}+\beta_\leftarrow e^{-ikj},
& j_1\leq j\leq j_2,\\[2mm]
e^{-ikj}+r_R(k)e^{ikj}, & j>j_2.
\end{cases}
\label{eq:right_ansatz}
\end{equation}
Repeating the same matching procedure gives
\begin{equation}
u_e^{(\leftarrow)}
=
\frac{V_-^*(k)}{D(k)},
\label{eq:ue_right}
\end{equation}
and hence
\begin{equation}
r_R(k) = -\frac{iV_+(k)V_-^*(k)}{v_g(k)D(k)},\quad t_R(k) = 1-\frac{i|V_-(k)|^2}{v_g(k)D(k)}.
\end{equation}
Thus the complete two-port scattering matrix can be obtained as follows
\begin{equation}
S(k)
=
\begin{bmatrix}
-\dfrac{iV_-(k)V_+^*(k)}{v_g(k)D(k)}
&
1-\dfrac{i|V_-(k)|^2}{v_g(k)D(k)}
\\[4mm]
1-\dfrac{i|V_+(k)|^2}{v_g(k)D(k)}
&
-\dfrac{iV_+(k)V_-^*(k)}{v_g(k)D(k)}
\end{bmatrix}.
\label{eq:S_final}
\end{equation}

\section{High-Order Double-Sided Output Zeros}
\label{Sec:III}
In this section, we investigate CPA under two-sided coherent excitation. We first identify the incident channel that is completely absorbed at a given wave vector and then determine the additional conditions under which the corresponding scattering zero becomes higher order.

\subsection{Coherent perfect absorption}
We consider coherent excitation from both ports, for which simultaneous suppression of the two outgoing channels corresponds to CPA. The coherent two-sided input can be characterized by $\boldsymbol{a}_\eta=1/\sqrt{1+|\eta|^2}(1,\eta)^T$, where $\eta=|\eta|e^{i\varphi}$. The corresponding output amplitudes $\boldsymbol{b}_\eta=[b_L(k),b_R(k)]^T$ can be obtained as follows
\begin{equation}\label{eq_b}
    \boldsymbol{b}_\eta=S(k)\boldsymbol{a}_\eta=\frac{1}{\sqrt{1+|\eta|^2}D(k)}
    \begin{bmatrix}
        N_L(k)\\
        N_R(k)
    \end{bmatrix},
\end{equation}
where 
\begin{equation}\label{eq_NLNR}
    \begin{aligned}
        N_L(k)&=\eta D(k)-\frac{iV_-(k)Y(k)}{v_g(k)},\\
        N_R(k)&=D(k)-\frac{iV_+(k)Y(k)}{v_g(k)},
    \end{aligned}
\end{equation}
with $Y(k)=V_+^*(k)+\eta V_-^*(k)$.

The CPA at $k=k_0$ requires $S(k_0)\boldsymbol{a}_\eta=0$. A nontrivial solution of $\boldsymbol{a}_\eta$ exists only when $\det [S(k_0)]=0$. For analytical convenience, we redefine the two coupling sites symmetrically about the coordinate origin as $j_{1}=-L/2$ and $j_{2}=L/2$. Using the scattering matrix derived in Eq. (\ref{eq:S_final}), we find $\det [S(k_0)]=-\Omega(k_0)/D(k_0)$, where $\Omega(k_0)=\omega(k_0)-\omega_a-\Delta_g(k_0)+i\left[\gamma-\Gamma_g(k_0)\right]/2$. The nontrivial scattering-zero condition can be recast as $\Omega(k_0)=0$ and $D(k_0)\neq 0$. Consequently, the real-part equality gives the resonant frequency matching condition
\begin{equation}\label{eq_1CPA1}
\omega(k_0)=\omega_a+\Delta_g(k_0),
\end{equation}
while the imaginary-part equality provides the dissipative matching condition
\begin{equation}\label{eq_1CPA2}
    \gamma=\Gamma_g(k_0).
\end{equation}
$\Gamma_g(k_0)$ characterizes the external radiative coupling between the giant atom and the cavity array modes. The same coupling channel governs both excitation of the atom by the array and radiation from the atom back into the array.

To achieve bandwidth-enhanced perfect absorption, we derive the parameter conditions for realizing different orders of CPA under different coupling-point separations and coupling phases in the following subsections, and reveal the physical mechanism where nonlocal interference phases and coupling phases cooperatively regulate the first-order and high-order scattering interference zeros. Following the definition of high-order CPA, an $n$th order CPA at $k=k_0$ at both ports requires
\begin{equation}
\left.\frac{d^s b_L(k)}{dk^s}\right|_{k=k_0}=0,\quad 
\left.\frac{d^s b_R(k)}{dk^s}\right|_{k=k_0}=0,
\end{equation}
with $s=0,1,\ldots,n-1$. Provided that $D(k_0)\neq 0$, an $n$th-order zero of $b_{\alpha}(k)$ with $\alpha=L,R$ at $k=k_{0}$ can be determined by the following conditions
\begin{equation}
    N^{(s)}_{\alpha}(k_0)=0,
    \quad
    s=0,1,\ldots,n-1,
    \label{eq:nth_zero_numerator}
\end{equation}
where the superscript $(s)$ denotes differentiation with respect to the wave vector $k$ and
\begin{equation}
    N^{(n)}_{\alpha}(k_0)\neq0.
    \label{eq:nth_zero_nonzero_derivative}
\end{equation}
Taylor expansion gives
\begin{equation}
    b_{\alpha}(k)
    =
    \frac{N^{(n)}_{\alpha}(k_0)}
    {n!\sqrt{1+|\eta|^2}D(k_0)}
    (\delta k)^n
    +
    O\left[(\delta k)^{n+1}\right],
    \label{eq:nth_zero_amplitude_expansion}
\end{equation}
where $\delta k=k-k_0$. Consequently, the leading term of the output intensity obeys 
\begin{equation}
    |b_{\alpha}(k)|^2
    \propto
    |\delta k|^{2n}.
    \label{eq:nth_zero_intensity_scaling}
\end{equation}
Thus, first-, second-, and third-order amplitude zeros produce
quadratic, quartic, and sixth-power intensity scalings, respectively.

We define the normalized total output intensity as
\begin{equation}
P_{\mathrm{out}}(k)=|b_L(k)|^2+|b_R(k)|^2.
\label{eq:Pout}
\end{equation}
Therefore, CPA at $k=k_0$ is equivalently characterized by $P_{\mathrm{out}}(k_0)=0$, which requires the simultaneous vanishing of both output amplitudes, $b_L(k_0)=b_R(k_0)=0$.

\subsection{First-order CPA}


\begin{table*}[t]
\caption{Classification of scattering zeros at $k_0=\pi/2$ under different lengths $L=4m-r$. Here $r=0,1,2,3$. The dark-state parameter values for which
$V_+(k_0)=V_-(k_0)=0$ are excluded. Other parameters are set as $j_1=-L/2$ and $j_2=L/2$.}
\centering
\renewcommand{\arraystretch}{1.4}
\setlength{\tabcolsep}{1mm}
\resizebox{\textwidth}{!}{%
\begin{tabular}{l
  r @{$\;=\;$} l
  r @{$\;=\;$} l
  r @{$\;\neq\;$} l
  r @{$\;=\;$} l
  r @{$\;=\;$} l
  r @{$\;=\;$} l
  r @{$\;=\;$} l
  c}
\hline\hline
$L$ & \multicolumn{8}{c}{Generic CPA conditions} & \multicolumn{6}{c}{Additional high-order conditions} & Highest order \\
\hline
\noalign{\vspace{0.5mm}}

$4m-3$
& $\displaystyle \omega_c-\omega_a$ & $\displaystyle \frac{g^2}{J}\cos\Delta\phi$,
& $\displaystyle \gamma$ & $\displaystyle \frac{2g^2}{J}$,
& $\displaystyle \Delta\phi$ & $\displaystyle \pm \frac{\pi}{2}$,
& $\displaystyle \eta_0$ & $\displaystyle \tan\left(\frac{\pi}{4}-\frac{\Delta\phi}{2}\right)$
& \multicolumn{6}{c}{---}
& 1st \\[1mm]

$4m-2$
& $\displaystyle \omega_c-\omega_a$ & $\displaystyle 0$,
& $\displaystyle \gamma$ & $\displaystyle \frac{2g^2}{J}(1-\cos\Delta\phi)$,
& $\displaystyle \Delta\phi$ & $\displaystyle 0$,
& $\displaystyle \eta_0$ & $\displaystyle -1$
& $\displaystyle \Delta\phi$ & $\displaystyle \pi$,
& $\displaystyle \gamma$ & $\displaystyle \frac{4g^2}{J}$,
& $\displaystyle J^2$ & $\displaystyle (2m-1)g^2$
& 2nd \\[1mm]

$4m-1$
& $\displaystyle \omega_c-\omega_a$ & $\displaystyle -\frac{g^2}{J}\cos\Delta\phi$,
& $\displaystyle \gamma$ & $\displaystyle \frac{2g^2}{J}$,
& $\displaystyle \Delta\phi$ & $\displaystyle \pm \frac{\pi}{2}$,
& $\displaystyle \eta_0$ & $\displaystyle \tan\left(\frac{\pi}{4}+\frac{\Delta\phi}{2}\right)$
& \multicolumn{6}{c}{---}
& 1st \\[1mm]

$4m$
& $\displaystyle \omega_c-\omega_a$ & $\displaystyle 0$,
& $\displaystyle \gamma$ & $\displaystyle \frac{2g^2}{J}(1+\cos\Delta\phi)$,
& $\displaystyle \Delta\phi$ & $\displaystyle \pi$,
& $\displaystyle \eta_0$ & $\displaystyle 1$
& $\displaystyle \Delta\phi$ & $\displaystyle 0$,
& $\displaystyle \gamma$ & $\displaystyle \frac{4g^2}{J}$,
& $\displaystyle J^2$ & $\displaystyle 2mg^2$
& 2nd \\[1.5mm]
\hline\hline
\end{tabular}%
}
\label{Tab1}
\end{table*}

First-order perfect absorption at $k=k_0$ requires the condition $N_L(k_0)=N_R(k_0)=0$, from which we can derive the relative complex amplitudes between the two incident waves
\begin{equation}\label{eq_eta}
\eta_0=\frac{V_-(k_0)}{V_+(k_0)}=\frac{\cos [(k_0L+\Delta\phi)/2]}{\cos [(k_0L-\Delta\phi)/2]}.
\end{equation}
This expression is valid under $V_-(k_0)V_+(k_0)\neq 0$. In this case, we obtain $N_L(k_0)=\eta_0 N_R(k_0)$. 

In the following analysis, we focus on the case $k_0=\pi/2$. At the band-center momentum $k_0$, increasing the coupling-point separation by one site changes the propagation phase $k_0L$ by $\pi/2$. The four possible values of $L$ modulo $4$ therefore generate four distinct interference configurations. For $L=4m-3$ ($m=1,2,3,\dots$), the first-order CPA conditions can be simplified to
\begin{equation}
    \omega_c-\omega_a=\frac{g^2\cos \Delta\phi}{J},\quad \gamma=\frac{2g^2}{J},\quad \eta_0=\tan \left(\frac{\pi}{4}-\frac{\Delta\phi}{2}\right).
\end{equation}
For $L=4m-1$, the corresponding conditions are
\begin{equation}
    \omega_c-\omega_a=-\frac{g^2\cos \Delta\phi}{J},\quad \gamma=\frac{2g^2}{J},\quad \eta_0=\tan \left(\frac{\pi}{4}+\frac{\Delta\phi}{2}\right).
\end{equation}
For the two cases discussed above, the phase values $\Delta\phi=\pm \pi/2$ give either $\eta_0=0$ or $|\eta_0|\rightarrow\infty$. The absorbing channel then collapses onto a single input port. These points describe single-sided perfect-absorption, but they are excluded from the present discussion of two-sided CPA.

For $L=4m-2$, the conditions are
\begin{equation}
    \omega_c=\omega_a,\quad \gamma= \frac{2g^2}{J}(1-\cos \Delta\phi),\quad \eta_0=-1.
\end{equation}
At this point, $\Delta\phi=0$ gives $\Gamma_g(k_0)=0$ and $V_+(k_0)=V_-(k_0)=0$, indicating that the giant atom becomes dark to the cavity array, which implies that CPA cannot be achieved. 
For $L=4m$, the conditions become
\begin{equation}
    \omega_c=\omega_a,\quad \gamma= \frac{2g^2}{J}(1+\cos \Delta\phi),\quad \eta_0=1,
\end{equation}
and the same radiation-decoupling condition occurs at $\Delta\phi = \pi$.

The conditions derived above ensure that the selected scattering channel vanishes at the single momentum $k_0$. However, they do not constrain how rapidly the output increases when the incident momentum deviates from $k_0$. To obtain a higher-order CPA, the same input vector $\boldsymbol{a}_{\eta_0}$ must be kept fixed and the leading momentum dependence of its output must also be eliminated.

\subsection{Second-order CPA}

\begin{figure*}
\centering
\includegraphics[width=1.97\columnwidth]{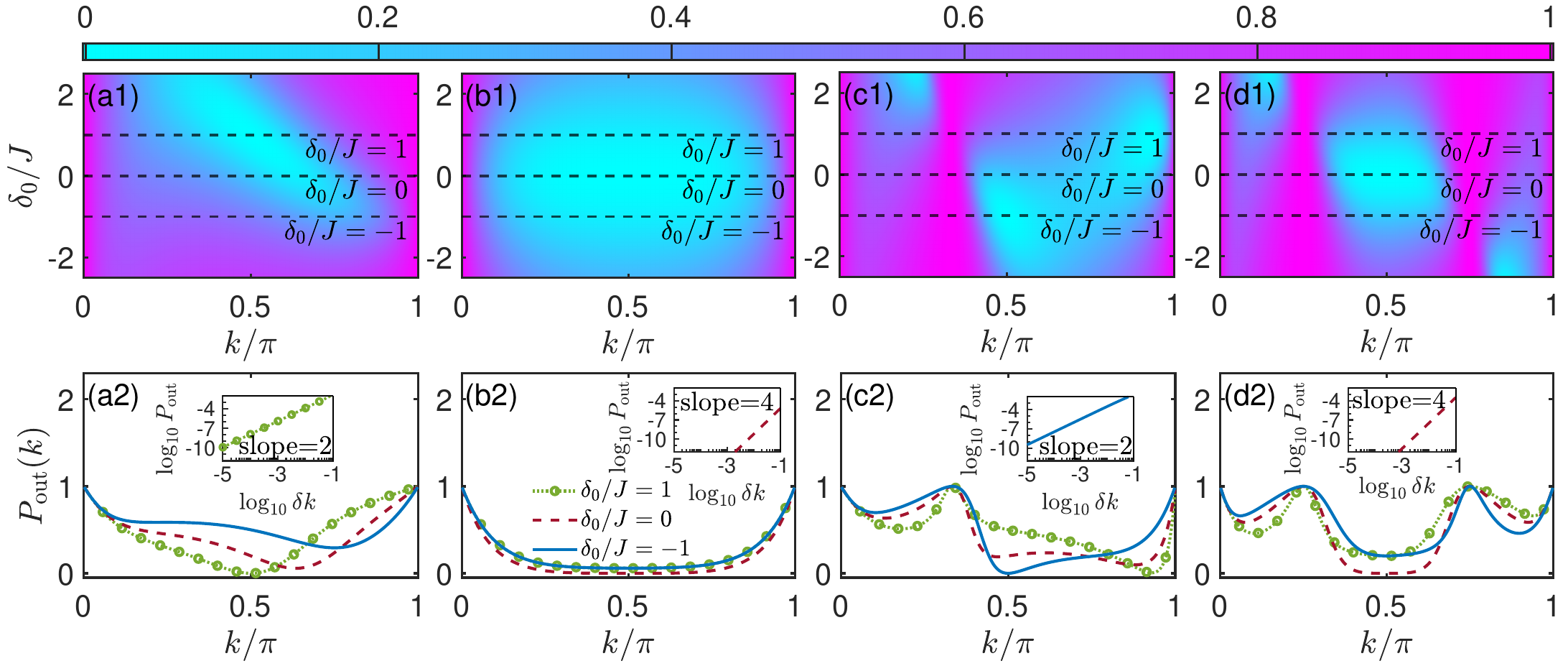}
\caption{Normalized total output intensity $P_{\mathrm{out}}(k)$ under
two-sided coherent excitation as a function of various system parameters. (a1)--(d1) Density plots as a function of wave vector $k$ and detuning $\delta_0$. Here $\delta_0=\omega_c-\omega_a$.
(a2)--(d2) The cuts indicated by the black dashed
lines. The inset displays the corresponding
double-logarithmic plots near $k_0=\pi/2$.
The four columns correspond to
(a) $L=1$, $\Delta\phi=0$, $\eta=1$, $g=J$, and
$\gamma=2J$;
(b) $L=2$, $\Delta\phi=\pi$, $\eta=-1$, $g=J$, and
$\gamma=4J$;
(c) $L=3$, $\Delta\phi=0$, $\eta=1$, $g=J$, and
$\gamma=2J$; and
(d) $L=4$, $\Delta\phi=0$, $\eta=1$,
$g=J/\sqrt{2}$, and $\gamma=2J$.}
\label{Fig.2}
\end{figure*}

Achieving second-order CPA necessitates $\det[S(k_0)]=0$ and the additional conditions $N_{L}'(k_0)=0$ and $N_{R}'(k_0)=0$ for left and right ports. To expose the physical content of the derivative conditions, we define $F(k)=V_-(k)/V_+(k)$. By using $F(k)$, Eqs.~\eqref{eq_NLNR} can be written as
\begin{equation}\label{eq_NLNR_2}
    \begin{aligned}
        N_L(k)&=\eta_0\Omega(k)+\frac{i|V_+(k)|^2}{v_g(k)}\left[\eta_0-F(k)\right],\\
        N_R(k)&=\Omega(k) + \frac{iV_+(k)V_-^*(k)}{v_g(k)}\left[F(k)-\eta_0\right].
    \end{aligned}
\end{equation}
These expressions separate two independent contributions. The function $\Omega(k)$ determines the scattering eigenvalue zero,
whereas $F(k)$ determines the momentum dependence of the absorbing input channel. 

Considering Eqs.~\eqref{eq_NLNR_2} and 
$F(k_0)=\eta_0$, we obtain the following additional conditions from the second‑order CPA requirements
\begin{equation}
    F'(k_{0})=0, \quad \Omega'(k_{0})=0.
\end{equation}
The condition $F'(k_0)=0$ prevents the absorbing channel from
rotating to first order in $\delta k$, while
$\Omega'(k_0)=0$ eliminates the linear variation of the associated scattering eigenvalue. 

Differentiating with respect to $k$ gives
\begin{equation}
    F'(k_{0})
    =
    -\frac{L\sin\Delta\phi}
    {2\cos^2\left[(k_{0}L-\Delta\phi)/2\right]}.
    \label{eq:R_prime_explicit}
\end{equation}
For a nontrivial two-sided CPA point with $V_+(k_0)\neq 0$,
Eq.~\eqref{eq:R_prime_explicit} requires $\sin\Delta\phi=0$, which gives $\Delta\phi=0$ and $\Delta\phi=\pi$. The condition $\Omega'(\pi/2)=0$ can be reduced to the following conditions
\begin{equation}\label{eq:Omega_prime}
    \begin{aligned}
       2J^2-g^2L\cos\Delta\phi\cos\left(\frac{\pi L}{2}\right)&=0,\\
       g^2L\cos\Delta\phi\sin\left(\frac{\pi L}{2}\right)&=0.
    \end{aligned}
\end{equation}
The second equation in Eqs. \eqref{eq:Omega_prime} implies $\sin(\pi L/2)=0$ and hence $L$ must be even. Therefore, odd coupling-point
separations support only first-order CPA at the band center. 

For the case $L=4m-2$, we can verify the relation $\cos(\pi L/2)=-1$. Then the first equation in Eqs. \eqref{eq:Omega_prime} gives $2J^{2}+Lg^2 \cos\Delta \phi=0$. Since $J^2$, $g^2$, and $L$ are positive, this equation requires $\cos\Delta\phi=-1$, or equivalently, $\Delta\phi=\pi$ and $\eta_0=-1$. Then the condition can be recast as $J^{2}=(2m-1)g^{2}$. Under these parameter choices, the eigenvalues and eigenvectors of the scattering matrix read $\lambda_{-1}^+(k_0)=1$, $\boldsymbol{a}_{-1}^+=1/\sqrt{2}(1,1)^T$ and $\lambda_{-1}^-(k_0)=0$, $\boldsymbol{a}_{-1}^-=1/\sqrt{2}(1,-1)^T$.

For $L=4m$, we have $\cos(\pi L/2)=1$. The first equation in Eqs. \eqref{eq:Omega_prime} then gives $2J^{2}-Lg^2 \cos\Delta \phi=0$, which requires $\cos\Delta\phi=1$, or equivalently, $\Delta\phi=0$ and $\eta_0=1$. Then the condition can be recast as $J^{2}=2m g^{2}$. Under these conditions, the eigenvalues of the scattering matrix are given by $\lambda_1^+(k_0)=0$, $\boldsymbol{a}_{1}^+=1/\sqrt{2}(1,1)^T$ and $\lambda_1^-(k_0)=-1$, $\boldsymbol{a}_{1}^-=1/\sqrt{2}(1,-1)^T$. Table~\ref{Tab1} summarizes the full set of parameter constraints for achieving CPA. At all families of first‑ and second-order CPA points, we have $D(\pi/2)=4ig^2/J\neq 0$. The first‑ and second-order zeros are therefore genuine scattering zeros and do not coincide with a scattering pole.

These results further establish that the second-order CPA does not originate from an EP. At $k=k_0$, the two scattering eigenvalues remain nondegenerate, and their eigenvectors remain orthogonal. The second-order CPA therefore corresponds to a double zero of a single scattering eigenvalue as a function of momentum, rather than to the coalescence of two eigenvalues and eigenvectors.

To provide a unified characterization of the two outgoing channels, Fig.~\ref{Fig.2} presents the normalized total output intensity $P_{\mathrm{out}}(k)$ as a function of wave vector $k$ and detuning $\delta_0$, with $\delta_0=\omega_c-\omega_a$. Because the incident vector is normalized, $P_{\mathrm{out}}(k)=1$ corresponds to complete outgoing transport, whereas $P_{\mathrm{out}}(k)=0$ indicates simultaneous suppression of both output ports and hence CPA.
For the separations $L=1$ and $L=3$, the CPA occurs at $\delta_0/J=1$ and $-1$, respectively. The corresponding double-logarithmic plots have slope $2$, showing that the output intensity increases quadratically with $\delta k$ and that these points are conventional first-order CPA zeros. In contrast, for the separations $L=2$ and $L=4$, the properly tuned coupling strengths and loss rates eliminate the linear momentum dependence of both output amplitudes. The total output intensity exhibits a quartic line shape, as evidenced by the slope $4$ in Figs.~\ref{Fig.2}(b2) and \ref{Fig.2}(d2). These flatter zeros demonstrate the enhanced momentum bandwidth associated with second-order CPA.

Under the second‑order conditions, we obtain 
\begin{equation}
    \Omega''(k_0)=i\frac{g^2}{J}(L^2-2).
\end{equation}
For all even integers $L\ge 2$ considered in this work, $L^2-2\neq 0$. Consequently, $\Omega''(k_0)\neq 0$, which implies that third‑order CPA cannot be realized within the present symmetric two-coupling-point model at $k_0=\pi/2$.

\section{High-Order Single-Sided Output Zeros}
\label{Sec:IV}
In Sec.~\ref{Sec:III}, we studied scattering-eigenvalue
zeros under two-sided coherent excitation. We now consider
single-sided excitation and investigate zeros of individual
scattering-matrix elements. In particular, we focus on the
reflection-amplitude zero and transmission-amplitude zero for a wave incident from the left port.

For left incidence, the input vector is $\boldsymbol{a}_L=(1,0)^T$ or equivalently $\eta=0$. Substituting $\eta=0$ into Eq. \eqref{eq_b}, one can obtain the simplified output amplitudes as follows
\begin{equation}
    b_L(k)=r_L(k)=\frac{N_L(k)}{D(k)},
    \quad
    b_R(k)=t_L(k)=\frac{N_R(k)}{D(k)},
    \label{eq:single_port_outputs}
\end{equation}
where
\begin{equation}
\label{eq:single_port_numerators}
\begin{aligned}
    N_{L}(k)=&-i \frac{g^2}{J\sin k}(\cos kL + \cos\Delta\phi),\\
    N_{R}(k)=&\omega(k)-\omega_a-\Delta_g(k)+i\left(\frac{\gamma}{2} - \frac{g^2\sin kL\sin\Delta\phi}{J\sin k}\right).
\end{aligned}
\end{equation}

\subsection{Zero reflection}

To realize the first-order zero reflection, the constraints $N_L(k_0)=0$ and $D(k_0)\neq 0$ must be satisfied, which reduces to
\begin{equation}\label{eq_0r}
    V_-(k_0)V_+^*(k_0)=0.
\end{equation}
This equation yields two independent solution branches, either $V_+(k_0)=0$ or $V_-(k_0)=0$, both of which correspond to zero reflection.

For the former case, the atomic excitation amplitude $u_e^{(\rightarrow)}=0$, which means the atom remains unexcited. Accordingly, perfect transmission $b_R=t_L(k_0)=1$ can be achieved even when $\gamma\neq0$ for the giant-atom waveguide system. This corresponds to dark-state transmission, which means the incident wave interferes destructively at the two coupling positions, preventing the giant atom from absorbing any energy. For the latter case, by contrast, the atom remains excited with $u_e^{(\rightarrow)}\neq0$. Nevertheless, interference completely cancels the radiation field propagating toward the left, yielding a vanishing leftward radiation amplitude of the atom and thus $r_L=0$. If both conditions hold simultaneously, the system becomes fully transparent.

In the case of $j_{1}=-L/2$ and $j_{2}=L/2$, we derive from Eq. (\ref{eq_0r}) that
\begin{equation}\label{eq_pt}
    \quad k_0L=\pm\Delta\phi+(2q+1)\pi, \quad q\in\mathbb{Z}.
\end{equation}
For $L=4m-3$ and $L=4m-1$, it reduces to $\Delta\phi=\pm \pi/2$, for which only one of the collective coupling amplitudes vanishes, yielding a first-order reflection zero. In the case $V_-(k_0)=0$, if the conditions given by Eqs.~\eqref{eq_1CPA1} and \eqref{eq_1CPA2} are satisfied, we obtain $b_R(k_0)=0$, corresponding to single‑port perfect absorption, as discussed earlier. By contrast, for $L=4m-2$, $\Delta\phi=0$ and $L=4m$, $\Delta\phi=\pi$, both collective coupling amplitudes vanish simultaneously, which gives rise to second-order output zeros. This second-order zero is thus a dark-state output zero, which corresponds to the giant atom being decoupled from the cavity array. These radiative-decoupling solutions are excluded from the following classification of nontrivial reflection zeros.

Interestingly, the simultaneous dark condition is also closely related to the formation of a bound state in the continuum (BIC) \cite{Zhang_2025,p67j-1r3b,l1fq-gbbl}. At this point, the radiative decay rate vanishes, $\Gamma_{g}(k_{0})=0$, and for the symmetric two-coupling-point configuration considered here,
the interference-induced frequency shift also satisfies $\Delta_{g}(k_{0})=0$. If the additional conditions
\begin{equation}
    \omega_{a}=\omega(k_{0}),\quad \gamma=0,
\end{equation}
are imposed, the dressed denominator becomes
$D(k_{0})=0$. The corresponding atomic--photonic excitation is
therefore resonant with the propagating band while being completely
decoupled from both propagation channels, which constitutes a BIC.
We emphasize, however, that this BIC point is distinct from the
regular reflection zeros considered above, for which
$D(k_{0})\neq0$. Thus, $V_{+}(k_{0})=V_{-}(k_{0})=0$ represents a
radiative-decoupling condition in general and becomes a BIC only
when the resonance and lossless conditions are simultaneously
satisfied.

\begin{figure}
\centering
\includegraphics[width=0.95\columnwidth]{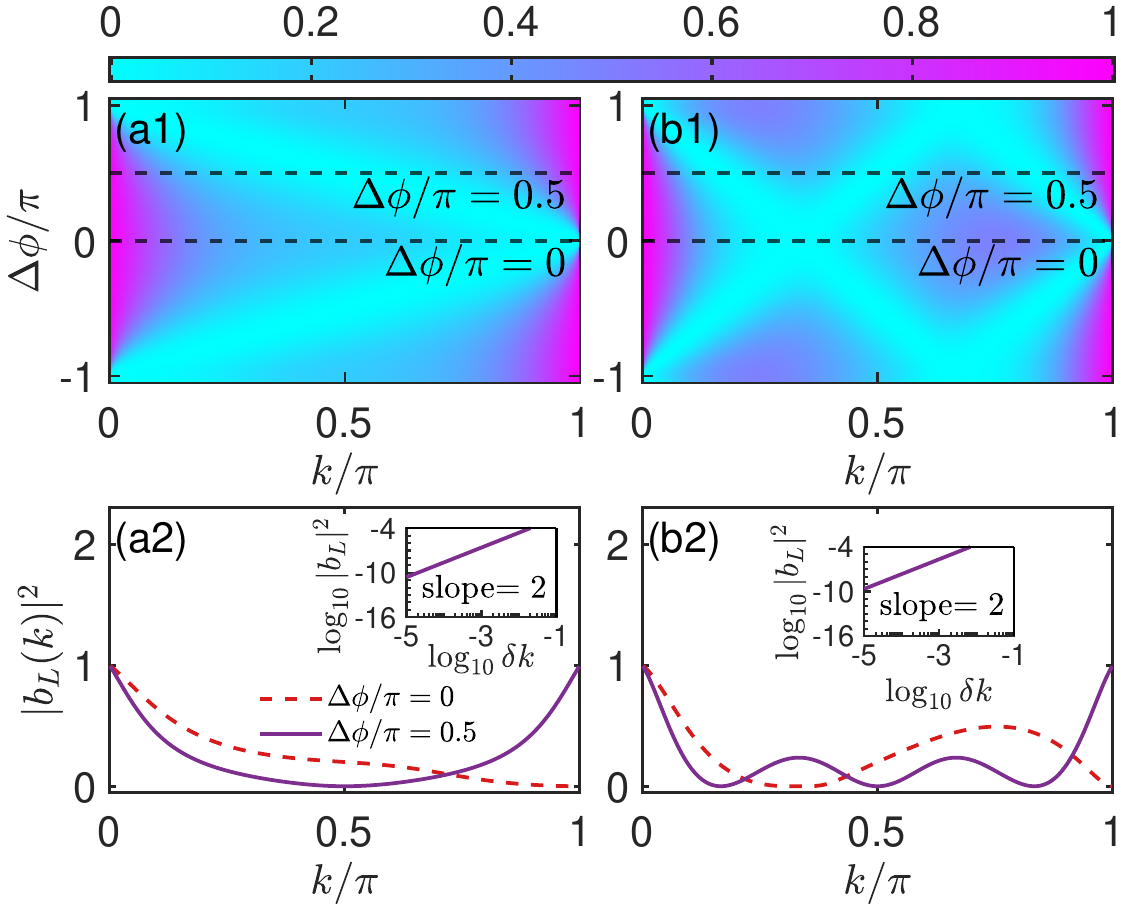}
\caption{Left-port reflection intensity $|b_L(k)|^2$ under single-sided incidence $(\eta=0)$ as a function of various system parameters. (a1)--(b1) Density plots as a function of wave vector $k$ and $\Delta\phi$ for $L=1$ and $3$, respectively. (a2)--(b2) The cuts indicated by the black dashed lines. The inset displays the corresponding double-logarithmic plots near $k_0=\pi/2$. The common parameters are $\omega_c=\omega_a$, $g=J$, and $\gamma=2J$.}
\label{Fig.3}
\end{figure}

Figure~\ref{Fig.3} provides a direct visualization of the reflection-zero conditions derived above. Figures~\ref{Fig.3}(a1) and \ref{Fig.3}(b1) show the left-port reflection intensity $|b_L(k)|^2$ as a function of the incident wave vector $k$ and the coupling-phase difference $\Delta\phi$ for $L=1$ and $L=3$, respectively. In particular, at the band-center momentum $k_0=\pi/2$, the choices $\Delta\phi=-\pi/2$ for $L=1$ and $\Delta\phi=\pi/2$ for $L=3$ yield $|b_L(k_0)|^2=0$, whereas the reference cuts at $\Delta\phi=0$ do not exhibit a reflection zero at $k_0$. Figures~\ref{Fig.3}(a2) and \ref{Fig.3}(b2) further illustrate the strong suppression of the reflected field at the target momentum. More importantly, the double-logarithmic plots exhibit a slope of $2$, confirming that these zeros are first-order zeros of the reflection amplitude.

\subsection{Zero transmission}

\subsubsection{First-order zero transmission}


\begin{table*}[t]
\caption{Classification of transmission zeros at $k_0=\pi/2$ for left incidence under different lengths $L=4m-r$. Here $r=0,1,2,3$. Other parameters are set as $j_1=-L/2$, $j_2=L/2$ and $\eta=0$.}
\centering
\renewcommand{\arraystretch}{1.4}
\setlength{\tabcolsep}{3mm}
\resizebox{\textwidth}{!}{%
\begin{tabular}{l
  r @{$\;=\;$} l @{\quad}
  r @{$\;=\;$} l @{\quad}
  l
  r @{$\;=\;$} l @{\quad}
  r @{$\;=\;$} l
  c}
\hline\hline
$L$ & \multicolumn{5}{c}{Generic zero-transmission conditions} & \multicolumn{4}{c}{Additional high-order conditions} & Highest order \\
\hline
$4m-3$
& $\displaystyle \omega_c-\omega_a$ & $\displaystyle \frac{g^2}{J}\cos\Delta\phi$,
& $\displaystyle \gamma$ & $\displaystyle \frac{2g^2}{J}\sin\Delta\phi$,
& $\displaystyle 0\leq\Delta\phi\leq \pi$
& \multicolumn{4}{c}{---}
& 1st \\[1mm]

$4m-2$
& $\displaystyle \omega_c-\omega_a$ & $\displaystyle 0$,
& $\displaystyle \gamma$ & $\displaystyle 0$,
& $\displaystyle \Delta\phi\neq 0$
& $\displaystyle \Delta\phi$ & $\displaystyle \pi$,
& $\displaystyle J^2$ & $\displaystyle (2m-1)g^2$
& $\begin{cases}
b_R(k)=0,\quad &m=1\\
\mathrm{3rd}, \quad &m \geq 2
\end{cases}$ \\[1mm]

$4m-1$
& $\displaystyle \omega_c-\omega_a$ & $\displaystyle -\frac{g^2}{J}\cos\Delta\phi$,
& $\displaystyle \gamma$ & $\displaystyle -\frac{2g^2}{J}\sin\Delta\phi$,
& $\displaystyle -\pi\leq\Delta\phi\leq 0$
& \multicolumn{4}{c}{---}
& 1st \\[1mm]

$4m$
& $\displaystyle \omega_c-\omega_a$ & $\displaystyle 0$,
& $\displaystyle \gamma$ & $\displaystyle 0$,
& $\displaystyle \Delta\phi\neq \pi$
& $\displaystyle \Delta\phi$ & $\displaystyle 0$,
& $\displaystyle J^2$ & $\displaystyle 2mg^2$
& 3rd \\[1.5mm]
\hline\hline
\end{tabular}%
}
\label{Tab2}
\end{table*}

Zero transmission occurs when the directly transmitted field is exactly canceled by the field radiated by the giant atom into the right-going channel. In a lossless system, energy conservation then implies that zero transmission is equivalent to perfect reflection. For a wave incident from the left, the condition for first-order zero transmission takes the form $N_R(k_0)=0$ when $D(k_0)\neq 0$, which leads to
\begin{equation}\label{eq_pr}
        \omega(k_0)=\omega_a+\Delta_g(k_0),\quad
        \gamma=\frac{2g^2\sin\Delta\phi\sin k_0L}{J\sin k_0}.
\end{equation}
These conditions hold for arbitrary wave vector $k_0\in(0,\pi)$ and arbitrary positive integer coupling-point separation $L$.

For the coupling-point separation $L=4m-3$, it can be simplified to
\begin{equation}
    \omega_c-\omega_a=\frac{g^2\cos\Delta\phi}{J}, \quad 
    \gamma=\frac{2g^2\sin\Delta\phi}{J},\quad 0\leq\Delta\phi\leq\pi.
\end{equation}
For $L=4m-1$, the conditions become
\begin{equation}
    \omega_c-\omega_a=-\frac{g^2\cos\Delta\phi}{J}, \quad 
    \gamma=-\frac{2g^2\sin\Delta\phi}{J},\quad -\pi\leq\Delta\phi\leq 0.
\end{equation}
The above formulas reveal that both the cavity-atom detuning and the dissipative rate are modulated by the phase difference $\Delta\phi$. In contrast, when $L=4m-2$ and $L=4m$, the conditions reduce to $\omega_c=\omega_a$, $\gamma=0$, which gives rise to perfect reflection.
Physically, this result indicates that first-order zero transmission for this case can only be realized when the atomic frequency resonates with the cavity mode and the dissipation rate of the giant atom vanishes. Although the resonance and loss conditions do not explicitly depend on $\Delta\phi$, the coupling phase cannot be completely arbitrary. The dark-state value $\Delta\phi=0$ for $L=4m-2$ and $\Delta\phi=\pi$ for $L=4m$ must be excluded because the system approaches perfect transmission $b_R(k_0)=1$ rather than zero transmission.

\subsubsection{High-order zero transmission}

We next discuss whether the first-order transmission zeros above can be promoted to higher order. An $n$th-order transmission zero requires
\begin{equation}
N_R^{(s)}(k_0)=0,
\quad
s=0,1,\ldots,n-1,
\label{eq:ZT_order_general}
\end{equation}
while $N_R^{(n)}(k_0)\neq0$ and $D(k_0)\neq 0$. At $k_0=\pi/2$, one has
\begin{equation}
N_R'(k_0)= 2J - \frac{Lg^2}{J} e^{i\Delta\phi}\cos\left(\frac{\pi L}{2}\right).
\label{eq:NRprime_bandcenter}
\end{equation}
Equation~(\ref{eq:NRprime_bandcenter}) reveals a fundamental distinction between odd and even coupling-point separations. For odd $L$, $\cos(\pi L/2)=0$, so that
\begin{equation}
N_R'(k_0)=2J\neq0.
\end{equation}
Consequently, no additional choice of the coupling phase, coupling strength, or dissipation can eliminate the linear momentum dependence. All such transmission zeros at the band center are therefore strictly first order. For even $L$, by contrast,
$\cos(\pi L/2)=\pm1$, and the linear term can be canceled.
For $L=4m$, Eq.~(\ref{eq:NRprime_bandcenter}) gives
\begin{equation}
\Delta\phi=0,
\quad
J^2=2mg^2,
\label{eq:ZT_high_4m}
\end{equation}
whereas for $L=4m-2$ it gives
\begin{equation}
\Delta\phi=\pi,
\quad
J^2=(2m-1)g^2.
\label{eq:ZT_high_4m2}
\end{equation}
Together with the first-order conditions, these conditions
eliminate both $N_R(k_0)$ and $N_R'(k_0)$. The transmission numerator can be written as
\begin{equation}
N_R(k_0+\delta k) = 2J\sin \delta k - ce^{i\Delta\phi} \frac{g^2\sin(L\delta k)}{J\cos \delta k},
\label{eq:NR_even_q}
\end{equation}
where $c=1$ for $L=4m$, and $c=-1$ for $L=4m-2$. Both terms in Eq.~(\ref{eq:NR_even_q}) are odd functions of $\delta k$. The Taylor expansion of $N_R(k_0+\delta k)$ therefore contains only odd powers of $\delta k$, implying $N_R''(k_0)=0$ identically for the even-$L$ transmission-zero family. Hence, once the linear term is removed, there is no isolated second-order transmission zero: the zero is automatically promoted to at least third order.

\begin{figure*}
\centering
\includegraphics[width=1.97\columnwidth]{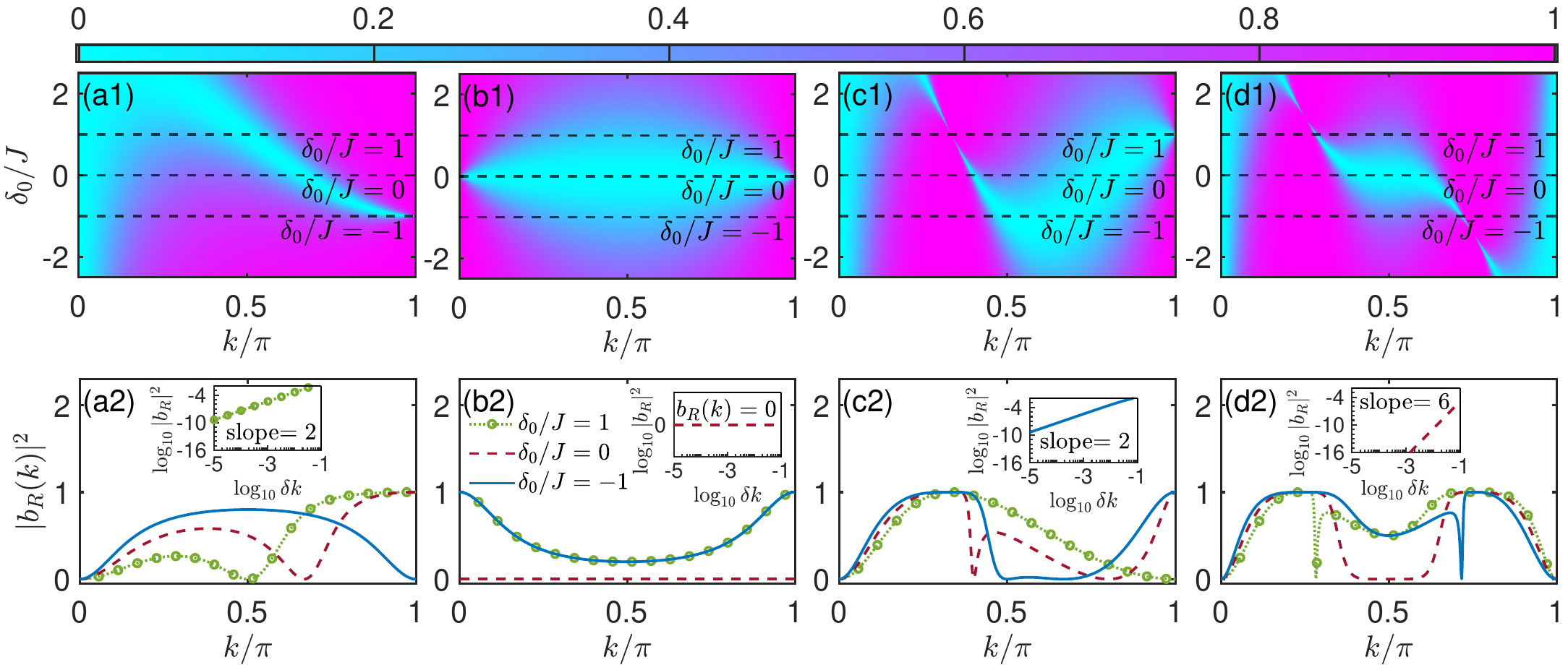}
\caption{Right-port transmission intensity
$|b_R(k)|^2=|t_L(k)|^2$ under single-sided incidence
$(\eta=0)$ as a function of various system parameters. (a1)--(d1) Density plots as a function of wave vector $k$ and detuning $\delta_0$. Here $\delta_0=\omega_c-\omega_a$. (a2)--(d2) The cuts along the black dashed lines. The inset displays the corresponding double-logarithmic plots near $k_0=\pi/2$. The four columns correspond to
(a) $L=1$, $\Delta\phi=0$, $g=J$, and $\gamma=0$;
(b) $L=2$, $\Delta\phi=\pi$, $g=J$, and $\gamma=0$;
(c) $L=3$, $\Delta\phi=0$, $g=J$, and $\gamma=0$; and
(d) $L=4$, $\Delta\phi=0$, $g=J/\sqrt{2}$, and $\gamma=0$.}
\label{Fig.4}
\end{figure*}

Substituting the high-order conditions Eqs.~\eqref{eq:ZT_high_4m} or
\eqref{eq:ZT_high_4m2} into
Eq.~\eqref{eq:NR_even_q} yields
\begin{equation}
N_R(k_0+\delta k)
=
\frac{J(L^2-4)}{3}\delta k^3
+
\mathcal{O}(\delta k^5).
\label{eq:NR_cubic_general}
\end{equation}
For every even separation $L>2$, the cubic coefficient is
nonzero. The leading term of the corresponding output intensity therefore scales as $|b_R(k_0+\delta k)|^2 \propto |\delta k|^6$. Thus, all admissible higher-order transmission zeros with even $L>2$ are exactly third-order.

The case $L=2$ is exceptional because the cubic coefficient
in Eq.~(\ref{eq:NR_cubic_general}) also vanishes. Under the parameter conditions for second‑order zero transmission, the transmission numerator becomes
\begin{equation}
N_R(k)=2\cos k\left(\frac{g^2}{J}-J\right).
\end{equation}
Therefore, for $g=J$, we have $N_R(k)\equiv 0$, and consequently $b_R(k)\equiv0$.
This momentum-independent transmission zero is not caused by dark-state decoupling. For $j_1=-1$, $j_2=1$, and $\Delta\phi=\pi$, the collective coupling amplitudes are
\begin{equation}
V_+(k)=2ig\sin k,\quad
V_-(k)=-2ig\sin k,
\end{equation}
which are finite throughout the propagating band $0<k<\pi$. The giant atom therefore remains coupled to both propagation directions. Instead, the field radiated by the giant atom into the forward channel exactly cancels the directly transmitted field for every momentum over the entire propagation band.

Table \ref{Tab2} collects the necessary parameter constraints for different orders of zero transmission for different coupling-point separations $L=4m-r$, where $r=0,1,2,3$, $j_1=-L/2$, $j_2=L/2$ and $\eta=0$. 
Figure~\ref{Fig.4} summarizes the transmission spectra for the four representative coupling-point separations. For $L=1$ and $L=3$, the transmission intensity vanishes at $\delta_0/J=1$ and $-1$, respectively. The slopes of $2$ in Figs.~\ref{Fig.4}(a2) and \ref{Fig.4}(c2) show that these are first-order transmission zeros.

A qualitatively different behavior occurs for $L=2$. Under the conditions $\omega_c=\omega_a$, $\Delta\phi=\pi$, and $g=J$, $b_R(k)$ vanishes identically over the entire propagating band, as shown in Figs.~\ref{Fig.4}(b1) and ~\ref{Fig.4}(b2). This momentum-independent zero produces exact bandwidth-enhanced perfect reflection.
For $L=4$, choosing $\Delta\phi=0$ and $g=J/\sqrt{2}$ suppresses the first- and second-order momentum variations of the transmission amplitude. The output intensity is dominated by the sixth-order term, as demonstrated by the slope $6$ in Fig.~\ref{Fig.4}(d2). This behavior confirms the third-order transmission zero predicted in Table~\ref{Tab2}.

\section{Wave-Packet Dynamics}
\label{Sec:V}

To verify our steady-state scattering predictions, we perform time-evolution simulations of a coherent superposition of single-photon wave packets. The initial excitation is constructed as a superposition of two oppositely incident Gaussian wave packets \cite{PhysRevB.74.205120}, whose explicit form reads
\begin{equation}
    \ket{\psi(0)} = \frac{1}{\sqrt{1+|\eta|^2}}(\ket{\psi_{L}} + \eta \ket{\psi_{R}}),
\end{equation}
where
\begin{equation}\label{eq_wave}
    \ket{\psi_{\alpha}} = \frac{1}{\sqrt[4]{\pi\sigma^2}} \sum_j e^{-\frac{(j-j_{\alpha})^2}{2\sigma^2}} e^{\pm ik_c (j-j_{\alpha})}\ket{j},
\end{equation}
for $\sigma\gg1$. Here, $\ket{\psi_{\alpha}}$ is a wave packet centered at $j_\alpha$, with central momentum $+k_c$ and $-k_c$ and relative amplitude ratio $\eta$, where $j$ denotes the lattice site index and $\sigma$ controls its spatial width. The momentum-space width of the packet is proportional to $1/\sigma$, consistent with the Heisenberg uncertainty relation. The quantum state of the finite coupled-cavity
array is evolved according to the Hamiltonian in Eq.~(\ref{eq_H})
\begin{equation}
|\psi(t)\rangle
=
e^{-iHt}|\psi(0)\rangle.
\label{eq:TD_evolution}
\end{equation}

\begin{figure*}
\centering
\includegraphics[width=1.95\columnwidth]{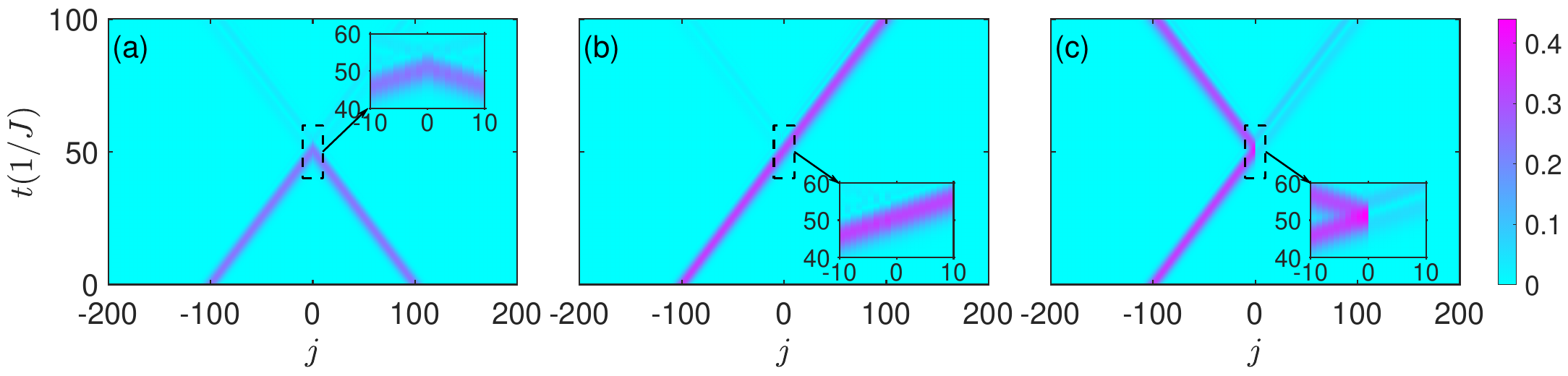}
\caption{Wave-packet dynamics for three representative scattering processes with coupling-point separation $L=1$. The color scale denotes the cavity field amplitude $|u_j(t)|$. The incident Gaussian wave packets have central momentum $k_c=k_0=\pi/2$ and spatial width $\sigma=5$. 
(a) First-order CPA under two-sided excitation with $j_L=-201/2$, $j_R=201/2$, $\Delta\phi=0$, $\eta=1$, $g=J$, $\omega_a=\omega_c-J$, and $\gamma=2J$.
(b) Zero reflection under left-sided incidence with $j_L=-201/2$,
$\Delta\phi=-\pi/2$, $g=J$, $\omega_a=\omega_c$, and $\gamma=2J$.
(c) Zero transmission under left-sided incidence with $j_L=-201/2$,
$\Delta\phi=0$, $g=J$, $\omega_a=\omega_c-J$, and $\gamma=0$, for which the incident wave packet is nearly perfectly reflected.}
\label{Fig.5}
\end{figure*}

Figure \ref{Fig.5} illustrates the dynamical evolution of scattering wave packets $|u_j(t)|$ with wave-packet width $\sigma=5$ under distinct scattering regimes for $L=1$. Figure \ref{Fig.5}(a) depicts the CPA realized by two counterpropagating Gaussian wave packets initially centered at $j_L=-201/2$ and $j_R=201/2$. Counterpropagating wave-packet components with equal incident weights propagate toward the scattering center without reflection. When the two wave-packet components reach the scattering region, they are almost completely absorbed, with only negligible residual fields observed in the reflection and transmission channels. This phenomenon verifies the CPA effect. Figure \ref{Fig.5}(b) shows the dynamics of a single incident wave packet centered at $j_L=-201/2$ under the zero-reflection condition. The light incident from the left input port propagates toward the scattering center without reflection. Upon reaching the position $j=0$, most of the light energy transmits through the scattering region. The reflected probability is strongly suppressed, which verifies the zero-reflection transport behavior of the system. Figure \ref{Fig.5}(c) presents the zero-transmission dynamics of another single wave packet with identical width, whose center is located at $j_L=-201/2$. The light incident from the left input port propagates to the scattering center at $j=0$ and is almost totally reflected. For Figs. \ref{Fig.5}(a), (b) and (c), finite spectral components away from $k_0$ generate small residual outputs. This phenomenon originates from the momentum-space broadening of Gaussian wave packets, which contain components detuned from the resonant wave vector $k_0=\pi/2$ and thus lead to faint residual outgoing signals.

To quantitatively characterize the influence of the finite
momentum width of the incident wave packet, we define the
residual output probability after the scattering process. The corresponding numerical residuals are evaluated directly
from the real-space wave function as
\begin{equation}\label{eq:numerical_residuals}
\mathcal{R}_{\rm CPA}^{\rm num}=\sum_{|j|>L/2}P_j,\quad
\mathcal{R}_{\rm ZR}^{\rm num}=\sum_{j<-L/2}P_j,\quad
\mathcal{R}_{\rm ZT}^{\rm num}=\sum_{j>L/2}P_j,
\end{equation}
with $P_j=|u_j(t_f)|^2$. Here $t_f$ denotes the time after the scattering process is completed but before any boundary‑reflected wave returns to the scattering region. In our numerical simulations, we take the total lattice size $N=400$ and choose the evaluation time $t_f=120/J$. This choice of $t_f$ is sufficiently long for the scattering process to finish, yet well before any wave reflected from the lattice boundaries can propagate back to the scattering region, so that $\mathcal{R}_\alpha$ only accounts for the direct scattering output without contamination from boundary‑reflected echoes. For CPA, both
outgoing channels are required to vanish and the residual is
therefore evaluated by summing over the two outgoing regions outside the scattering domain. For zero reflection and zero transmission, by contrast, only the corresponding output port is included.

The residual is affected not only by the spectral width but also by a possible detuning of the central momentum from $k_0$. In the large-system and narrow-momentum-distribution limit,
the discrete momentum spectrum becomes quasicontinuous and
the Gaussian wave packet can be treated analytically within the continuum limit. Defining $\delta k_c=k_c-k_0$, one has $\delta k=(k-k_c)+\delta k_c$. Using the Gaussian‑moment relations, the residual for a sufficiently broad real‑space Gaussian wave packet takes the following asymptotic form (derived in Appendix~\ref{APPENDIX:A})
\begin{equation}
    \mathcal{R}_{\xi}^{(n)}
    \simeq
    C_{\xi}^{(n)}
    \sum_{\ell=0}^{n}
    \binom{2n}{2\ell}
    (\delta k_c)^{2n-2\ell}
    \frac{(2\ell-1)!!}{2^\ell\sigma^{2\ell}},\quad \xi\in\{\mathrm{CPA},\mathrm{ZR},\mathrm{ZT}\}.
    \label{eq:general_residual_main}
\end{equation}
Here the double factorial is defined as $(2\ell-1)!! = 1\cdot 3\cdot 5\cdots (2\ell-1)$ for positive integers $\ell$, with the convention $(-1)!!=1$. The coefficients are $C_{\rm{CPA}}^{(n)}= |\beta_{L,n}|^2+ |\beta_{R,n}|^2$, $C_{\rm{ZR}}^{(n)}=|\beta_{L,n}|^2$, $C_{\rm{ZT}}^{(n)}=|\beta_{R,n}|^2$ and $\beta_{\alpha,n} = b_\alpha^{(n)}(k_0)/n!$. Equation~\eqref{eq:general_residual_main} thus unifies two distinct imperfections: the terms containing
$\sigma^{-1}$ originate from the finite spectral width of the wave
packet, whereas $\delta k_c$ characterizes a systematic mismatch
between the packet center and $k_0$.
The leading residuals for first-, second-, and third-order zeros become
\begin{equation}
\begin{aligned}
    &\mathcal{R}^{(1)}_\xi \simeq C_\xi^{(1)} \left[ (\delta k_c)^2 + \frac{1}{2\sigma^2} \right],\\
    &\mathcal{R}^{(2)}_\xi \simeq C_\xi^{(2)} \left[ (\delta k_c)^4 + \frac{3(\delta k_c)^2}{\sigma^2} + \frac{3}{4\sigma^4} \right],\\
    &\mathcal{R}^{(3)}_\xi \simeq C_\xi^{(3)} \left[ (\delta k_c)^6 + \frac{15(\delta k_c)^4}{2\sigma^2} + \frac{45(\delta k_c)^2}{4\sigma^4} + \frac{15}{8\sigma^6} \right].
\end{aligned}
\label{eq:R_kc_sigma}
\end{equation}

\begin{figure}
\centering
\includegraphics[width=0.95\columnwidth]{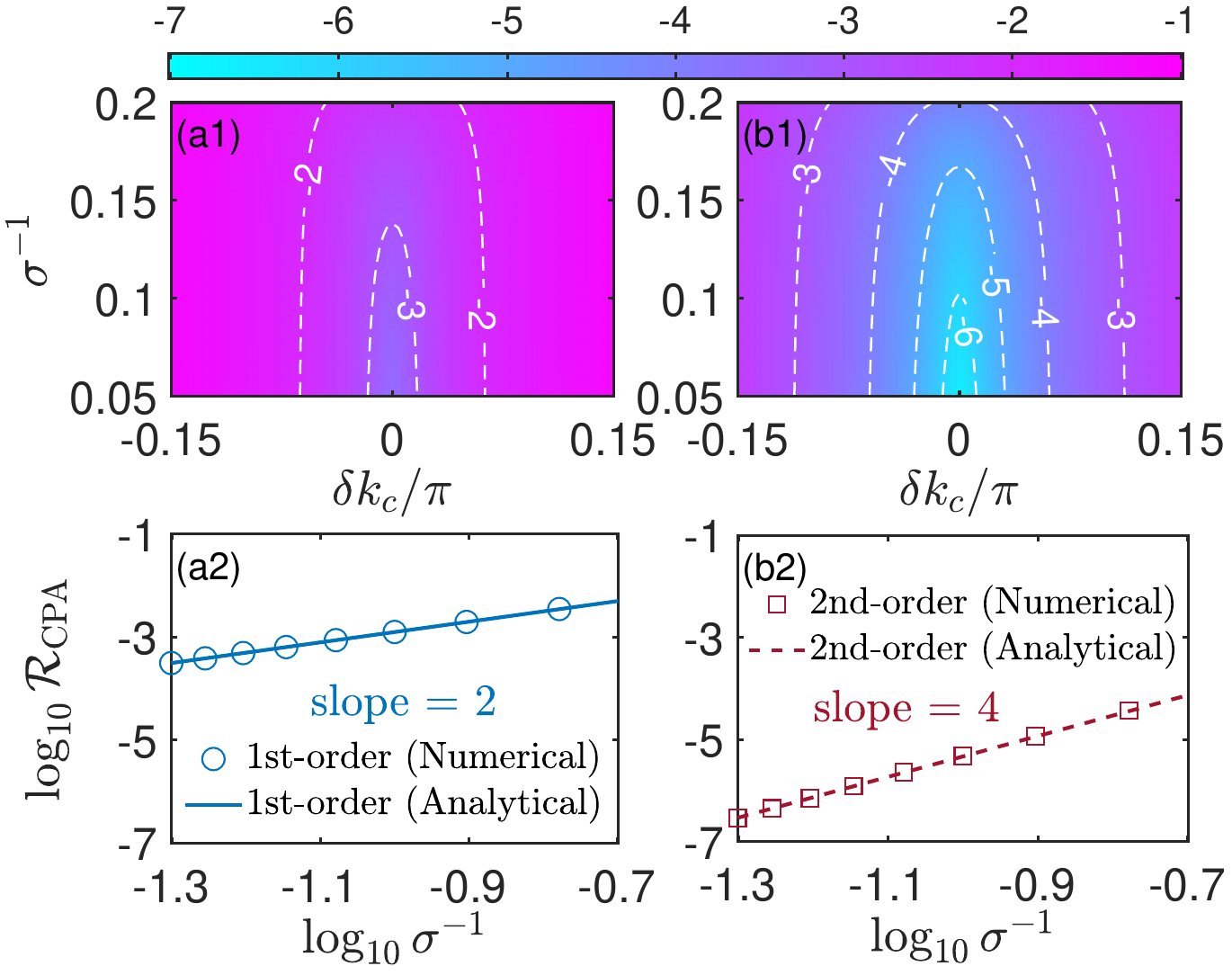}
\caption{Residual probabilities of finite-bandwidth Gaussian wave
packets centered at $k_c=k_0+\delta k_c$.
(a1) and (b1) Residual intensity $\log_{10}\mathcal{R}_{\rm{CPA}}$ as a function of $\sigma^{-1}$ and $\delta k_c/\pi$ computed from Eq.~(\ref{eq:numerical_residuals}) for first- and second-order CPA for $g=J/\sqrt{2}$, $\gamma=2J$ and $g=J$, $\gamma=4J$, respectively. White dashed lines denote contour lines of different residual values. 
(a2) and (b2) Residual intensity $\log_{10}\mathcal{R}_{\rm CPA}$ as a function of $\log_{10}\sigma^{-1}$ at $\delta k_c=0$, for first‑order and second‑order CPA, respectively.
Symbols are obtained from direct real-space time evolution
of a finite coupled-cavity array using Eq.~(\ref{eq:TD_evolution}), followed by the residual definition Eqs.~(\ref{eq:numerical_residuals}). Solid lines are the analytical large-$\sigma$ asymptotes Eq.~(\ref{eq:universal_sigma_scaling}) derived in Appendix~\ref{APPENDIX:A}. Other parameters are $L=2$, $\Delta\phi=\pi$, $\omega_c=\omega_a$, $\eta=-1$.}
\label{Fig.6}
\end{figure}

Thus, higher-order zeros suppress both the finite spectral width and the central-momentum detuning more efficiently. As indicated by Eqs.~(\ref{eq:R_kc_sigma}), increasing $\sigma$ is most effective when $k_c$ lies close to $k_0$. Away from the target momentum, the residual eventually becomes dominated by $\delta k_c$ instead of the finite spectral width.
This behavior can be visualized in Figs.~\ref{Fig.6}(a1) and \ref{Fig.6}(b1), which show the residual intensity for first‑order and second‑order CPA as functions of $\sigma^{-1}$ and the central‑momentum detuning $\delta k_c$, respectively.
These color maps display the residual probabilities obtained from direct real‑space wave‑packet numerical simulations, with white dashed contour lines marking equal‑residual boundaries.

When $k_c=k_0$, Eq.~\eqref{eq:general_residual_main} reduces to 
\begin{equation}
\mathcal{R}_{\xi}^{(n)} \simeq C_{\xi}^{(n)} \frac{(2n-1)!!}{2^n\sigma^{2n}},
\label{eq:universal_sigma_scaling}
\end{equation}
demonstrating the characteristic
$\mathcal{R}_{\xi}^{(n)}\propto\sigma^{-2n}$ suppression associated with an
$n$th-order amplitude zero. In this parameter regime, higher-order output zeros provide a direct suppression of the residual generated by the finite momentum bandwidth of the incident wave packet.

We compare the analytical results for CPA of different orders from Eq.~\eqref{eq:universal_sigma_scaling} with numerical evaluations of Eq.~\eqref{eq:numerical_residuals}.
Figures~\ref{Fig.6}(a2) and \ref{Fig.6}(b2) present the residual intensity versus the inverse wave-packet width $\sigma^{-1}$ for first‑ and second‑order CPA, where $\mathcal{R}_{\rm{CPA}}^{(1)}\simeq 1/8\sigma^2$ and $\mathcal{R}_{\rm{CPA}}^{(2)}\simeq 3/64\sigma^4$. On these double-logarithmic plots, the solid lines are the large‑$\sigma$ analytical asymptotes given by Eq.~\eqref{eq:universal_sigma_scaling}, with a slope of $2n$, corresponding to an $n$th‑order output zero.
The symbols correspond to numerical solutions obtained via real‑space time‑evolution simulations of the finite coupled‑cavity array.
The good agreement between them demonstrates that the finite-bandwidth residual provides a direct dynamical signature of the output-zero order.

\section{Experimental Feasibility}
\label{Sec:VI}

In this section, we briefly discuss the experimental feasibility of our proposed scheme, which can be realized with current experimental platforms. In a one-dimensional single-mode cavity array, individual cavities are coupled via evanescent fields, enabling photon hopping between adjacent cavities. In experiments, it can be realized with superconducting transmission line resonators \cite{RevModPhys.93.025005,science.1181918,Wallraff2004}. Giant-atom systems can be realized by coupling multiple small atoms to a waveguide at spatially separated positions \cite{Kannan2020}. Exploiting quantum interference arising from photon emission by an artificial giant molecule composed of two superconducting qubits, on-demand directional photon emission has been demonstrated experimentally \cite{Kannan2023}. Furthermore, local coupling phases can be introduced via Josephson-junction loops subject to external flux biasing \cite{Wang_2022,Chen2022}. Under current experimental conditions, the coupling strength $g_{1,2}/2\pi$ and inter-cavity nearest-neighbor coupling $J/2\pi$ can reach $10-100\ \text{MHz}$, while the resonator frequency $\omega_c/2\pi$ and giant-atom transition frequency $\omega_a/2\pi$ typically lie within the range of $1-10\ \text{GHz}$ \cite{PhysRevX.11.041043,PhysRevX.11.011015}. The pronounced separation of frequency justifies the validity of the rotating-wave approximation. Furthermore, superconducting artificial atoms can be engineered to achieve high coupling efficiencies to one-dimensional waveguide modes, where the guided-mode decay channel dominates over other dissipation channels \cite{Mirhosseini2019}. Our work focuses on a giant atom coupled to a coupled-cavity array, yet the relevant conclusions can be generalized to systems consisting of giant atoms interacting with LC-circuit resonators \cite{l1fq-gbbl}, acoustic waveguides \cite{Manenti2017}, or transmission-line waveguides \cite{PhysRevA.103.023710,Kannan2020}.

\section{Conclusion}
\label{Sec:VII}

In summary, we investigate CPA and single-photon scattering mediated by a two-site coupled giant atom in a one-dimensional coupled-cavity array. Using the wave-function matching method, we obtain the scattering matrix of the system and reveal that the spatial structure of the giant atom can introduce tunable nonlocal interference phases during scattering. These nonlocal interference phases, together with coupling phases, jointly determine the high-order output zeros in the absorption and transmission channels, offering a new degree of freedom for the manipulation of single-photon transport. The proposed system constitutes an effective scheme to realize tunable CPA as well as high-order transmission zeros in coupled-cavity array and waveguide quantum electrodynamics platforms. Moreover, the regulation mechanism based on separation-induced nonlocal interference of a giant atom can be further extended to systems with multiple coupling sites and multiple giant atoms. By adding more coupling sites and adjusting their respective coupling strengths, it is promising to achieve higher‑order CPA. This work provides feasible routes toward the realization of bandwidth-enhanced coherent perfect absorbers and single-photon switches and holds promise for applications in quantum photonic networks.

\section*{Acknowledgments}
G.W. was supported by the Quantum Science and Technology-National Science and Technology Major Project (No. 2023ZD0300700). H.L. was supported by the National Natural Science Foundation of China (Grants No. 12575011) and the Science and Technology Development Plan Project of Jilin Province, China (Grant No. 20240101321JC). H.S. was supported by Science and Technology Development Plan Project of Jilin Province (Grant No. 20250102007JC), and National Natural Science Foundation of China under Grant No. 12274064.

\appendix
\section{Residual intensity of three scattering processes}\label{APPENDIX:A}

The output zeros discussed in the previous sections are defined for monochromatic incident waves. A realistic wave packet, however, contains a finite range of momenta around its central wave vector. Consequently, even when the central wave vector coincides with the target output-zero wave vector $k_0$, spectral components away from $k_0$ may generate a finite residual output. We now establish a quantitative connection between the order of an output zero and the residual intensity of a finite-bandwidth wave packet.

We generalize the Gaussian wave packets in Eq.~(\ref{eq_wave}) to an arbitrary central wave vector $k_c$. For a spatial width $\sigma$ sufficiently larger than the lattice constant, the corresponding momentum envelope can be approximated by a continuous integral. The momentum‑space envelopes for the left‑ and right‑incident wave packets then read

\begin{equation}
\widetilde{A}_{\alpha}(k) = \mathcal{N}_k\exp\left[-\frac{\sigma^2}{2}(k-k_c)^2\right]\exp\left[\mp i(k-k_c)j_\alpha\right],
\label{eq:Gaussian_k}
\end{equation}
where $\mathcal{N}_k$ is the normalization constant, and $j_\alpha$ denotes the center position of the wave packet at the initial time. Here the upper and lower signs correspond to the left- and
right-incident wave packets, respectively.

For the CPA wave packets considered here, the initial centers are
chosen symmetrically with respect to the scattering region,
$j_R=-j_L=j_0$. Consequently, the two incident packets possess the same momentum-space envelope 
\begin{equation}
\widetilde{A}_{L}(k)
=
\widetilde{A}_{R}(k)
\equiv
\widetilde{A}_{\sigma}(k),
\end{equation}
where
\begin{equation}
\widetilde{A}_{\sigma}(k) = \mathcal{N}_k\exp\left[-\frac{\sigma^2}{2}(k-k_c)^2\right]\exp\left[i(k-k_c)j_0\right],
\end{equation}
For the two-sided coherent excitation considered in the CPA configuration, the momentum-space input can be written as
\begin{equation}
\widetilde{\boldsymbol{\Psi}}_{\rm in}(k) = \widetilde{A}_{\sigma}(k) \boldsymbol{a}_{\eta_0},
\label{eq:wavepacket_input_k}
\end{equation}
where $\boldsymbol{a}_{\eta_0}$ is the coherent input channel selected at the target wave vector $k_0$. Importantly, $\eta_0$ is kept fixed for all momentum components of the wave packet. After scattering, the momentum-space output is
\begin{equation}
\widetilde{\boldsymbol{\Psi}}_{\rm out}(k,t) = \widetilde{A}_{\sigma}(k) S(k)\boldsymbol{a}_{\eta_0} e^{-i\omega(k)t}.
\label{eq:wavepacket_output_k}
\end{equation}
Accordingly, the wave function in either output channel is
\begin{equation}
\widetilde{\psi}^{\rm out}_\alpha(k,t) = \widetilde{A}_{\sigma}(k) b_\alpha(k) e^{-i\omega(k)t},
\label{eq:output_channel_k}
\end{equation}
where $b_\alpha(k)$ is the corresponding monochromatic scattering amplitude.

To quantify the residual output, we define the integrated residual intensity in channel $\alpha$ as
\begin{equation}
\begin{aligned}
    \mathcal{R}_\alpha &\equiv \int_0^\pi dk \left| \widetilde{\psi}^{\rm out}_\alpha(k,t) \right|^2\\
    &= \int_0^\pi dk \left| \widetilde{A}_\sigma(k) \right|^2 |b_\alpha(k)|^2.
\end{aligned}
\label{eq:Rs_k}
\end{equation}
Near an $n$th-order amplitude zero at $k=k_0$, the corresponding output amplitude admits the local expansion
\begin{equation}
b_\alpha(k) = \beta_{\alpha,n}(k-k_0)^n + \mathcal{O}\!\left[(k-k_0)^{n+1}\right],
\label{eq:bs_local_expansion}
\end{equation}
with $\beta_{\alpha,n} = b_\alpha^{(n)}(k_0)/n!$.

For CPA, the total residual output is therefore $\mathcal{R}_{\rm CPA} = \mathcal{R}_L+\mathcal{R}_R$.
For the single-sided zero reflection and zero transmission, the same construction applies with the fixed input vector $\boldsymbol{a}_L=(1,0)^T$, where only the undesired output channel is retained. We thus obtain $\mathcal{R}_{\rm ZR} = \mathcal{R}_L$ and $\mathcal{R}_{\rm ZT} = \mathcal{R}_R$. For the CPA configurations considered here, the two output amplitudes possess the same zero order $n$. We then define the prefactors $C_{\rm{CPA}}^{(n)}= |\beta_{L,n}|^2+ |\beta_{R,n}|^2$, $C_{\rm{ZR}}^{(n)}=|\beta_{L,n}|^2$ and $C_{\rm{ZT}}^{(n)}=|\beta_{R,n}|^2$. Consequently, the three processes admit a unified form for $\xi\in\{\mathrm{CPA},\mathrm{ZR},\mathrm{ZT}\}$
\begin{equation}
    \mathcal{R}_{\xi}^{(n)}\simeq C_{\xi}^{(n)}\left\langle(k-k_0)^{2n}\right\rangle_\sigma.
\label{eq:R_moment}
\end{equation}
Here $\langle\cdot\rangle_\sigma$ denotes an average over the normalized momentum distribution $\left| \widetilde{A}_{\sigma}(k) \right|^2$ of the incident wave packet, defined as
\begin{equation}
    \langle f(k) \rangle_\sigma\equiv\int_{0}^{\pi} dk \bigl|\widetilde{A}_{\sigma}(k)\bigr|^2 f(k).
\label{eq:avg_def}
\end{equation}

We first allow the center of the incident wave packet to be slightly detuned from $k_0$, defined by $\delta k_c=k_c-k_0$.
We decompose $k-k_0=(k-k_c)+\delta k_c$ and evaluate the required Gaussian moments under the condition $\sigma^{-1}\ll \min(k_c,\pi-k_c)$. Under this condition, the Gaussian distribution is exponentially suppressed before reaching the band edges at $k=0$ and $k=\pi$. The integration interval may therefore be extended from $(0,\pi)$ to $(-\infty,\infty)$ with exponentially small corrections, yielding
\begin{equation}
    \left\langle(k-k_c)^{2\ell}\right\rangle_\sigma\simeq\frac{(2\ell-1)!!}{2^\ell\sigma^{2\ell}},\quad\left\langle(k-k_c)^{2\ell+1}\right\rangle_\sigma\simeq0,
\label{eq:Gaussian_moments}
\end{equation}
and we arrive at Eq.~\eqref{eq:general_residual_main} presented in the main text.

For a wave packet centered exactly at $k_0$, i.e., $\delta k_c=0$, only the term with $\ell=n$ in Eq.~(\ref{eq:general_residual_main}) remains and the residual reduces to Eq.~(\ref{eq:universal_sigma_scaling}). This result establishes the universal scaling $\mathcal{R}_{\xi}^{(n)}\propto\sigma^{-2n}$ for regular finite-order output zeros, showing that higher-order amplitude zeros suppress residual output caused by finite momentum bandwidth progressively more strongly.

\bibliography{ref}
\providecommand{\noopsort}[1]{}\providecommand{\singleletter}[1]{#1}%

\end{document}